\documentclass[aps,prl,superscriptaddress,floatfix,twocolumn,showpacs]{revtex4}

\usepackage{verbatim}

\usepackage{amsmath}
\usepackage{amsfonts}
\usepackage{amssymb}
\usepackage{graphicx}
\usepackage{bm}
\usepackage{color}
\usepackage{epsf}
\usepackage[hypertex]{hyperref}

\usepackage{psfrag}

\newcommand{\fV}{f_{3\,\rho}^V}
\newcommand{\fA}{f_{3\,\rho}^A}

\newcommand{\zV}{\zeta_{3}^V}
\newcommand{\zA}{\zeta_{3}^A}

\def\kb{$\underline{k}$}

\def\bea{\begin{eqnarray}}
\def\eea{\end{eqnarray}}
\def\beas{\begin{eqnarray*}}
\def\eeas{\end{eqnarray*}}
\def\beqas{\begin{eqnarray*}}
\def\eqas{\end{eqnarray*}}
\def\beq{\begin{equation}}
\def\eeq{\end{equation}}
\def\beqd{\begin{displaymath}}
\def\eeqd{\end{displaymath}}
\def\eqd{\end{displaymath}}

\def\slashchar#1{\setbox0=\hbox{$#1$}
   \dimen0=\wd0
   \setbox1=\hbox{/} \dimen1=\wd1
   \ifdim\dimen0>\dimen1
      \rlap{\hbox to \dimen0{\hfil/\hfil}}
      #1
   \else\begin{eqnarray}
      \rlap{\hbox to \dimen1{\hfil$#1$\hfil}}
      /
   \fi}

\begin{document}
\title
{On the description of exclusive processes beyond the leading twist approximation
  \\}

\author{ I.V.~Anikin}
\affiliation{Bogoliubov Laboratory of Theoretical Physics,
JINR, 141980 Dubna, Russia }
\author{D.~Yu.~Ivanov}
\affiliation{ Institute of Mathematics, 630090 Novosibirsk, Russia}
\author{ B.~Pire}
\affiliation{ CPhT, \'Ecole Polytechnique,
CNRS, F-91128 Palaiseau,     France }
\author{ L.~Szymanowski}
\affiliation{ So{\l}tan Institute for Nuclear Studies,
Ho\.za 69, 00-681 Warsaw, Poland}
\author{ S.~Wallon}
\affiliation{ LPT, Universit\'e d'Orsay, CNRS, 91404 Orsay, France \\ 
UPMC Univ. Paris 6, facult\'e de physique, 4 place Jussieu, 75252 Paris Cedex 05, France }


\begin{abstract}

\noindent
We describe  hard exclusive processes beyond the leading twist
approximation in a framework based on
the Taylor expansion of the amplitude around the dominant light-cone directions.
This naturally introduces an appropriate set of non-perturbative correlators whose number is minimalized after taking into account QCD equations of motion and  the invariance under rotation on the light-cone.
 We examplify this method at the twist 3 level and  show that the  coordinate and momentum space descriptions are fully equivalent.
\end{abstract}

\maketitle


Exclusive reactions have been the scene of much recent progress in the domain of perturbative
QCD physics. Collinear factorization of the amplitude for vector meson production
has been demonstrated
at the leading twist level \cite{fact}. In this case the meson field appears
through twist 2 distribution amplitudes (DAs) which are defined 
as the Fourier transforms of the matrix element of nonlocal two particle
operators $ {\cal O}(x)$, such as $\bar{\psi}(x) \,\Gamma  \psi(0)\,$
for fermionic partons, $\Gamma$ being an appropriate Dirac structure.
Generalizing this factorization procedure to the twist 3 case is mandatory
in a number of cases  (provided the value of the hard scale is sufficiently high to justify phenomenologically  the twist expansion), and in particular for the understanding of the
production of transversally polarized vector mesons \cite{MP}, since their twist 2
DAs are often decoupled because of their chiral-odd nature \cite{DGP}.

Following the strategy outlined in \cite{AT}, we here generalize the
approaches \cite{AT, EFP}
in order to get a thorough description of the exclusive processes.
In particular, we discuss the role of the light-cone direction  independency condition and
 fully explore its consequences  in the analysis of genuine higher twist effects, in particular beyond the Wandzura Wilczek approximation.  We show that the present
framework is equivalent
to the coordinate description developed in \cite{BalBraun,BB}.

\begin{figure}[h]
\psfrag{l}[cc][cc]{$\ell$}
\psfrag{lm}[cc][cc]{}
\psfrag{q}[cc][cc]{$\gamma^*$}
\psfrag{H}[cc][cc]{$H$}
\psfrag{S}[cc][cc]{$\Phi$}
\psfrag{Hg}[cc][cc]{$H_\mu$}
\psfrag{Sg}[cc][cc]{$\Phi^\mu$}
\psfrag{k}[cc][cc]{}
\psfrag{rmk}[cc][cc]{}
\psfrag{rho}[cc][cc]{$\rho$}
\begin{tabular}{cccc}
\includegraphics[width=3.8cm]{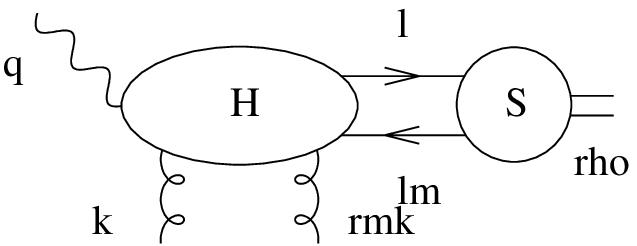}&\hspace{-.2cm}
\raisebox{.7cm}{+}&\hspace{-.1cm}\includegraphics[width=3.8cm]{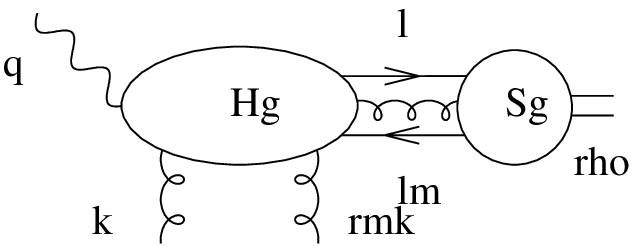}
&\hspace{-.2cm}\raisebox{3.5 \totalheight}{$+ \cdots$}
\end{tabular}
\caption{2- and 3-body correlators attached to a hard scattering amplitude in the  example of the $\gamma^* \to \rho$ impact factor.}
\label{fig:NonFactorized}
\end{figure}

In this paper we summarize only the main ideas of our developments, which are to a big extent valid for arbitrary exclusive processes. However, for illustration
of the main steps of the derivation, in particular of the 
implementation of the invariance under rotation on the 
light-cone of the physical amplitude, we will rely on the specific 
meson impact factor case  at twist 3, when needed.
This quantity is illustrated 
in Fig. \ref{fig:NonFactorized}, it enters  the description of 
transversly polarized vector mesons production 
at small $x$, i.e. at large energy,  in $\gamma^*\gamma^*$ and $\gamma^* p$ collisions.
For the aims of present paper it is enough to understand that 
it represents a part of the amplitudes of above 
mentioned processes which describes the transition of a 
virtual photon to a vector meson mediated by two virtual gluons 
being in the colorless state.
Full details of the derivations and explicit definition of 
impact factor are given in \cite{Us}. 

Thus, we start with the amplitude presented in Fig.\ref{fig:NonFactorized}, written as (with $\{d^4 \ell \}_2=d^4\ell_1\, d^4\ell_2$)
\begin{eqnarray}
\label{GenAmp}
\!{\cal A}=
\!\!\int \!d^4\ell \,\, {\rm tr} \biggl[ H(\ell) \Phi (\ell) \biggr] 
\!\!
+\!\!\int \!\{d^4 \ell \}_2\, {\rm tr}\biggl[
H_{\mu}(\ell) \Phi^\mu (\ell) \biggr] \!\!+\! ..
\end{eqnarray}
where $H$ and $H_{\mu}$ are the hard parts
with two parton legs and two parton and one gluon legs,
respectively. It is important to note that it is assumed that virtualities of $t-$channel gluons in Fig. \ref{fig:NonFactorized} are large, being of order of photon virtuality $Q^2$. Thus, all nonperturbative, soft momenta part of the subprocess described by the impact factor is included 
in the blobs $\Phi$ and $\Phi^\mu$.  
 In (\ref{GenAmp}), the soft parts are given by the
Fourier-transformed two or three body correlators.
  Inclusion of 4-body correlators would lead to twist  $\geq 4$ contributions, which are suppressed by additional powers of $1/Q$. Besides, here we consider the case of vanishing transverse momentum transfer in the $\gamma^*\to\rho$ transition and we neglect corrections of the order $m_\rho/Q$.

We have to parameterize momenta and polarization vector of the transversly polarized meson ($e$). 
In our subprocess $\gamma^*(q)\to \rho(p_\rho)$, $q^2=-Q^2$, $p_\rho^2=0$,
\begin{equation}
q_\mu=p_\mu-\frac{Q^2}{2}\tilde n_\mu\ , \quad p_{\rho\,\mu}=p_\mu \ .
\label{new1}
\end{equation}
Here $p$ ($\tilde n$) be a  ``plus" (``minus") light-cone vector, respectively (normalized as $p \cdot \tilde n=1$).
Such light-cone vectors can be easily constructed from the 4- vectors 
describing the kinematics of the process in question. 

The polarization vector of transversly polarized meson obeys
\begin{equation}
e \cdot p =0 \ , \quad e \cdot \tilde n =0 \ .
\label{new2}
\end{equation}
This vector coincides with the physical polarization vector in the infinite $\rho-$meson momentum frame.

The amplitude (\ref{GenAmp}) is not factorized yet because the hard and soft parts are related by
the 4-dimensional integration in the momentum space and by the summation over
the Dirac and/or Lorentz indices.
To factorize the amplitude, we  need to select  the dominant direction around  which
we intend to decompose our relevant momenta and the hard part.  For this purpose we choose another pair of 
light-cone vectors: $p$ and $n$ (normalized again as $p \cdot n=1$). 
It is very important to note that here we take the freedom to choose $n$ arbitrary, having no direct relationship with the parametrization of external momenta, $n\neq\tilde n$. 
We carry out an expansion of $\ell$ in the basis defined by $p$ and $n$:
\begin{eqnarray}
\label{k}
\ell_{i\, \mu} &=& y_i\,p_\mu  + (\ell_i\cdot p)\, n_{\mu} + \ell^\perp_{i\,\mu}
\end{eqnarray}
with
$y_i=\ell_i\cdot n \,,$
and rewrite the integration measure in (\ref{GenAmp}) as
$d^4 \ell_i \to d^4 \ell_i \, dy_i \, \delta(y_i-\ell\cdot n)\,.$
We also decompose the hard part
around the dominant ``plus" direction:
\begin{equation}
\label{expand}
\vspace{-.4cm}
H(\ell) = H(y p) + \frac{\partial H(\ell)}{\partial \ell_\alpha} \biggl|_{\ell=y p}\biggr. \,
(\ell-y\, p)_\alpha + \ldots\,,
\end{equation}
with $(\ell-y \, p)_\alpha \approx \ell^\perp_\alpha $ up to twist 3 accuracy, where $\alpha$ denotes transverse components  with respect to vectors $p$ and $n$.

The $l^\perp$ dependence of the hard part looks first as an excursion out of
the collinear framework.
The factorized expression is obtained as the result of an integration by parts
which replaces  $\ell^\perp_\alpha$ by $\partial^\perp_\alpha$ acting on
the soft correlator.
 This leads to new operators
${\cal O}^\perp$ which contain
transverse derivatives, such as $\bar \psi \, \partial^\perp \psi $,
and thus
to the necessity of considering additional correlators
$\Phi^\perp (l)$.
This procedure leads to the factorization of the amplitude in momentum
space. Factorization in the Dirac space can be achieved by
the Fierz decomposition. For example, in the case of two fermions,  one should project out the Dirac matrix
$\psi_\alpha(0) \bar \psi_\beta(z)$
 which appears in the soft part of the amplitude on the relevant $\Gamma$ matrices.
Thus,  the 
amplitude finally takes the simple factorized form:
\begin{eqnarray}
\label{GenAmpFac}
&&
{\cal A}=
\int\limits_{0}^{1} dy \,{\rm tr} \left[ H(y) \, \Gamma \right] \,
\Phi^{\Gamma} (y)\,
 \nonumber \\
&&
+\int\limits_{0}^{1} dy_1\, dy_2 \,{\rm tr} \left[ H^\mu(y_1,y_2)
\, \Gamma \right] \, \Phi^{\Gamma}_{\mu} (y_1,y_2) .
\end{eqnarray}
The first term of Eq. (\ref{GenAmpFac})   requires the introduction of the soft correlators with
 quark-antiquark
nonlocal operators. Generically \cite{EPS},
\begin{eqnarray}
\label{CorrelatorV}
&&\langle \rho(p)|\bar\psi(z)\gamma_{\mu} \psi(0)|0\rangle
\nonumber \\
&&\hspace*{1.2cm} \stackrel{{\cal F}_1}{=} m_\rho \, f_\rho \,[
\varphi_1(y)\, (e^*\cdot n)p_{\mu}+\varphi_3(y)\, e^{*\,T}_{\mu}]\,,
\\
\label{CorrelatorA}
&&\langle \rho(p)|
\bar\psi(z)\gamma_5\gamma_{\mu} \psi(0) |0\rangle
\nonumber \\
&&\hspace*{2cm}\stackrel{{\cal F}_1}{=}m_\rho \, f_\rho \,
i\varphi_A(y)\,
\varepsilon_{\mu\lambda\beta\delta}\,
e^{*\,T}_{\lambda}\, p_{\beta} \, n_{\delta}\,,
\end{eqnarray}
where
$\stackrel{{\cal F}_1}{=}$
denotes the Fourier transformation
$\int_{0}^{1}\, dy \,\text{exp}\left[iy\,p\cdot z\right]\,.$
The light-like separation $z=\lambda \, n$
involves an arbitrary light-like vector $n$ with $n \cdot p=1$
which already appeared in the Sudakov decomposition (\ref{k}).
The above correlators are defined in the axial light-like gauge $n \cdot A=0\,,$
which allows to get rid of  Wilson lines. Thus, $n$ plays a triple role: it first fixes the axial gauge, second it  defines the transverse momentum of partons 
forming the $\rho$ in the collinear expansion, and last it determines the notion $e^{T}$ of transverse polarization of the $\rho\,$ in the above equations.  Note that $e^{T}$ should not be confused with the physical polarization vector of transverse meson, defined by the conditions (\ref{new2}).

The second term of Eq. (\ref{GenAmpFac})  requires the introduction of
 matrix elements of quark-transverse gluon 
operators
\begin{eqnarray}
\label{Correlator3BodyV}
\langle \rho(p)|
\bar\psi(z_1)\gamma_{\mu}g A_{\alpha}^T(z_2) \psi(0) |0\rangle
&\stackrel{{\cal F}_2}{=}&
\\
&&\hspace{-3cm}m_\rho \,\fV\,
B(y_1,y_2)\, p_{\mu} e^{*T}_{\alpha}\,,\nonumber
\\
\label{Correlator3BodyA}
\langle \rho(p)|
\bar\psi(z_1)\gamma_5\gamma_{\mu} g A_{\alpha}^T(z_2) \psi(0) |0\rangle
&\stackrel{{\cal F}_2}{=}&
 \\
&&\hspace{-3cm}m_\rho \,\fA\,
i \,
D(y_1,y_2)\, p_{\mu}
\varepsilon_{\alpha\lambda\beta\delta}
\, e^{*T}_{\lambda} \, p_{\beta} \, n_{\delta}\,,\nonumber
\end{eqnarray}
where
$\int\limits_{0}^{1} dy_1 \,\int\limits_{0}^{1} dy_2 \,
\text{exp}\left[ iy_1\,p\cdot z_1+i(y_2-y_1)\,p\cdot z_2 \right]$
is denoted by $\stackrel{{\cal F}_2}{=}\,.$
Note that the positivity of the gluon light-cone momentum fraction
imposes that quark-gluon parameterizing functions have the form
\begin{eqnarray}
B/D(y_1,y_2) &\equiv& {\cal B}/{\cal D}(y_1,y_2; y_2-y_1) \, \theta(y_1 \leq y_2\leq 1)\,.\nonumber
\end{eqnarray}
Here, the light-cone fractions of the quark,
anti-quark and gluon are respectively $y_1$, $1-y_2$ and $y_2-y_1\,.$

Finally, the second term of (\ref{expand}) leads to the introduction of the correlators
\begin{eqnarray}
\label{CorrelatorDerV}
&&\hspace{-.5cm}\langle \rho(p)|
\bar\psi(z)\gamma_{\mu}
i\stackrel{\longleftrightarrow}
{\partial^\perp_{\alpha}} \psi(0)|0 \rangle
\stackrel{{\cal F}_1}{=}m_\rho \, f_\rho \,
\varphi_1^T(y) \, p_{\mu} e^{*\,T}_{\alpha} \, ,
\\
\label{CorrelatorDerA}
&&\langle \rho(p)| \bar\psi(z)\gamma_5\gamma_{\mu}
i\stackrel{\longleftrightarrow}
{\partial^\perp_{\alpha}} \psi(0) |0\rangle
\nonumber \\
&&\hspace*{1cm}
\stackrel{{\cal F}_1}{=}m_\rho \, f_\rho \,
i \, \varphi_A^T (y) \, p_{\mu} \,
\varepsilon_{\alpha\lambda\beta\delta}
\, e^{*\,T}_{\lambda} \, p_{\beta} \, n_{\delta}\,.
\end{eqnarray}

Let us recall the symmetry properties of DAs \cite{AT} coming from $C$
invariance (with $\bar{y} \equiv 1-y$):
\begin{eqnarray}
\label{sym2and3Body}
&&\hspace{-.0cm}\varphi_1(y)=\varphi_1(\bar{y})\,, \quad  \varphi_3(y)=\varphi_3(\bar{y})\,, \quad
\varphi_A(y)=-\varphi_A(\bar{y})\,,  \nonumber \\
&&\hspace{-.0cm}\varphi^T_1(y)=-\varphi^T_1(\bar{y}) \,, \quad \varphi^T_A(y)=\varphi^T_A(\bar{y}) \,, \nonumber\\
&&\hspace{-.0cm}{\cal B}(y_1,y_2;y_g)=-{\cal B}(\bar{y}_2,\bar{y}_1;y_g)\,, \nonumber \\
&&{\cal D}(y_1,y_2;y_g)={\cal D}(\bar{y}_2,\bar{y}_1;y_g) .
\end{eqnarray}

The correlators introduced above are not independent,
thanks to the QCD equations of motion for the field
operators entering them (see, for example, \cite{AT}).
In the simplest case of fermionic fields, they follow from the vanishing matrix elements
$\langle (i  
{\hat D}(0) \psi(0))_\alpha\, \bar \psi_\beta(z)\rangle = 0$ and
$\langle  \psi_\alpha(0)\, i 
({\hat D}(z)\bar \psi(z))_\beta \rangle
= 0\,$
due to the Dirac equation, then projected on different Fierz structure.
Denoting $\zeta_{3, \, \rho}^{V,A}=f_{3 \,\rho}^{V,A}/f_\rho$, we obtain
\begin{eqnarray}
\label{em_rho1}
 &&\bar{y}_1 \, \varphi_3(y_1) +  \bar{y}_1 \, \varphi_A(y_1)  +  \varphi_1^T(y_1)  +\varphi_A^T(y_1)\nonumber \\
&&=-\int\limits_{0}^{1} dy_2 \left[ \zV \, B(y_1,\, y_2) +\zA \, D(y_1,\, y_2) \right] \,,
\end{eqnarray}
\begin{eqnarray}
\label{em_rho2}
 && y_1 \, \varphi_3(y_1) -  y_1 \, \varphi_A(y_1)  -  \varphi_1^T(y_1)  +\varphi_A^T(y_1) \nonumber\\
&&=-\int\limits_{0}^{1} dy_2 \left[ -\zV \, B(y_2,\, y_1) +\zA\, D(y_2,\, y_1) \right] \,.
\end{eqnarray}
Note that Eq.(\ref{em_rho2}) can be obtained by the replacement $y_1 \to \bar{y}_1$ in (\ref{em_rho1}) and the use of symmetry properties (\ref{sym2and3Body}).

 We stress again that contrarily to
the light-cone vector $p$ related to the out-going meson momentum,
 the second light-cone vector $n$, required for the parametrization of the
correlators (\ref{CorrelatorV}-\ref{CorrelatorDerA}), 
is completely arbitrary and connected in no way to the momenta of 
particles of the process in question.
Alternatively, the notions of transverse momentum, of 
transverse polarization and of axial gauge, lead to a description of the 
factorization procedure which is non-covariant. One should thus impose, 
due to Lorentz invariance, that any
 physical observables do not depend on the specific choice of $n$,
e.g. the scattering amplitudes should be $n-$independent.
It turns out that this requirement leads 
to additional non-trivial constraints between 
the non-perturbative correlators entering the 
factorized amplitude. This $n-$independency property was crucial
for the understanding of inclusive structure functions properties
at
 the twist three level  \cite{EFP} and its relevance for  
some exclusive processes was pointed out in \cite{AT}. 
We show now that this  condition expressed at the level 
of the {\em full amplitude} of any process can be reduced to a set of conditions involving only the soft correlators. The strategy
relies on the power of the Ward identities
to relate firstly amplitudes  with different number of legs and secondly higher order coefficients in the Taylor expansion (\ref{expand}) to lower order ones.

At first sight 
the $n-$independency condition of the process amplitude could not lead to some process independent results.
Indeed,  
not only the parameterizations of the hadronic matrix elements but also the 
process dependent hard scattering coefficient $H$ depends (indirectly)
on $n$, since $n$ defines the direction of the collinear expansion, see eq. (\ref{expand}). 
With this respect we remind that using integration    
by parts in the $d^4l$ integral (\ref{GenAmp}) one
replaces  $\ell^\perp_\alpha$ by $\partial^\perp_\alpha$ acting on
the soft correlator. Therefore the $n-$ independency of the
process amplitude can be reduced entirely in considered approximation (at twist-3) to the 
consideration of the $n-$dependency of the soft correlators, see 
eqs. (\ref{CorrelatorA},\ref{CorrelatorV},\ref{CorrelatorDerV},\ref{CorrelatorDerA},\ref{Correlator3BodyV}
 ,\ref{Correlator3BodyA}), 
which parameterizations
depend on $n$ both explicitly and implicitly through the 
$n-$dependence of the polarization vector  $e^T_\perp$.

The light-cone vector $n$ satisfying $n\cdot p=1$  may be parametrised  by its 
transverse components $n_\perp$ defined with respect to the light-cone
vectors $p$ and $\tilde n$ fixed by the external kinematics (see eq. (\ref{new1}))
\begin{equation}
\label{1}
n^\alpha\,=\,-\frac{n_{\perp}^2}2 \, p^\alpha + \tilde n^\alpha + n_{\perp}^\alpha\,.
\end{equation}
Thus  the $n-$independence of ${\cal A}$
for an arbitrary fixed polarization vector $e$  
 is expressed by the condition \cite{RG}
\begin{equation}
\label{2}
\frac{d}{dn_{\perp}^{\mu}}\;{\cal A}=0 \ .
\end{equation}

On the other hand, the scattering amplitude ${\cal A}$ receives contributions from the vector correlators, which result into ${\cal A}^{vector}$, and from the axial vector correlators, which lead to ${\cal A}^{axial}$ part of ${\cal A}$.
Due to different parity properties of the vector and the axial-vector correlators, the condition  (\ref{2}) means effectively two separate conditions:
\begin{equation}
\label{3}
\frac{d}{dn_{\perp}^{\mu}}\;{\cal A}^{vector}=0\;
\end{equation}
and
\begin{equation}
\label{4}
\frac{d}{dn_{\perp}^{\mu}}\;{\cal A}^{axial}=0\;.
\end{equation}

The dependence of ${\cal A}$ on the vector $n_\perp$ is obtained through the dependence of
${\cal A}$
 on the full vector $n$. This dependence on $n$ is different in ${\cal A}^{axial}$ and in ${\cal A}^{vector}$ parts.

\noindent
The dependence of ${\cal A}^{axial}$ on the vector $n$
enters only through the expression ${\varepsilon}^{p\,n\,\beta\,\gamma}$ involving the contraction with the momentum $p$ and in which the indices $\beta$ and $\gamma$ are contracted with some other vectors. 
Thus the condition (\ref{4}) is equivalent to
\begin{eqnarray}
\label{nIndA}
\frac{\partial\,n^\alpha}{\partial\, n_{\perp}^{\mu}}\,\frac{\partial}{\partial\,n^\alpha}\;{\cal A}^{axial} &=&
[ -n_{\perp}^\mu p^\alpha + g_{\perp}^{ \alpha\,\mu}]\frac{\partial}{\partial\,n^\alpha}\;{\cal A}^{axial}\nonumber\\
&=&
\frac{\partial}{\partial\,n_\perp^\mu}{\cal A}^{axial}=0
\end{eqnarray}
where  we took into account the peculiar dependence of ${\cal A}^{axial}$ on $n$ discussed above. This will lead at the level of DAs to the equation
\begin{eqnarray}
\label{ninA}
&&\frac{d}{dy_1}\varphi_A^T(y_1)=\varphi_A(y_1)-\zA\int\limits_0^1\,\frac{dy_2}{y_2-y_1}
 \nonumber \\
&&\times \left(D(y_1,y_2)+D(y_2,y_1)
   \right) \,.
\end{eqnarray}
\noindent
The  polarization vector for transverse  $\rho$ which enters in the
parametrization of twist 3 correlators  is defined as
\begin{equation}
\label{epsilonRho}
e_{T}^\mu = e^\mu - p^\mu\, e\cdot n\,.
\end{equation}
It turns out, as one can check explicitely in this example, that the
dependence
 of ${\cal A}^{vector}$ on the vector $n$ enters only through
 the scalar product $e^*\cdot n$, and Eq.~(\ref{3}) can be written as
\begin{equation}
\label{nIndV}
\frac{d}{dn_{\perp}^{\mu}}\;{\cal A}^{vector}= e_T^{*\,\mu}\,\frac{\partial}{\partial\,(e^*\cdot n)}\;{\cal A}^{vector}=0\;,
\end{equation}
from which follows
\begin{eqnarray}
\label{ninV}
&&\frac{d}{dy_1}\varphi_1^T(y_1)=-\varphi_1(y_1)+\varphi_3(y_1)-\zV\int\limits_0^1\,\frac{dy_2}{y_2-y_1}
\nonumber\\
&&\times \left( B(y_1,y_2)+B(y_2,y_1) \right)\,.
\end{eqnarray}

\begin{figure}[tb]
\psfrag{yq}{{\tiny $\hspace{-.2cm}y_1$}}
\psfrag{dy}{\raisebox{.04cm}{\tiny $\hspace{-.3cm}y_2-y_1$}}
\psfrag{yb}{{\tiny $\hspace{-.2cm}1-y_2$}}
\hspace{-0.2cm}\raisebox{.9cm}{$(y_1-y_2)$}\  \includegraphics[width=2.cm]{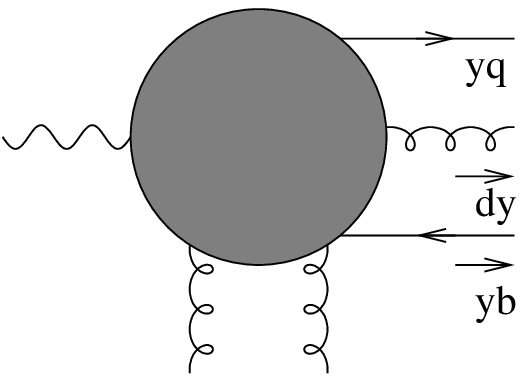}\hspace{.2cm} \raisebox{.8cm}{=}\,
\psfrag{yq}{{\tiny $\hspace{-.2cm}y_1$}}
\psfrag{yb}{{\tiny $\hspace{-.2cm}1-y_1$}}
\hspace{0.cm}\includegraphics[width=2.cm]{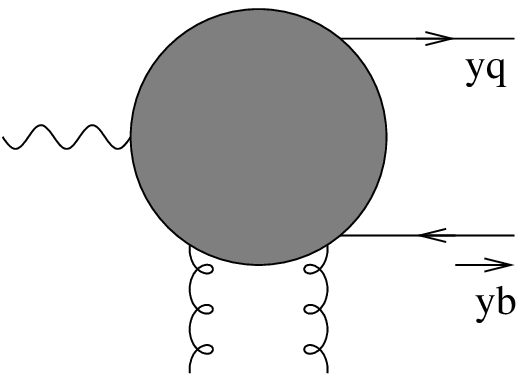}\hspace{0cm}\raisebox{.8cm}{-}\hspace{.1cm}
\psfrag{yq}{{\tiny $\hspace{-.2cm}y_2$}}
\psfrag{yb}{{\tiny $\hspace{-.3cm}1-y_2$}}
\includegraphics[width=2.cm]{RWardIF.eps}
\caption{Reduction of
 3-body correlators to 2-body correlators through Ward identity.}
\label{fig:WardIF}
\end{figure}

We now sketch  the derivation of Eq.(\ref{ninV}), which corresponds to the vector correlator contributions,
 in the peculiar case of the
$\gamma^* \to \rho$
impact factor, which is the building block of the
description of the $\gamma^* \, p \to \rho \, p$ process at large $s\,.$
The 3-body ($q \bar{q} g$) contribution  and the 2-body contribution involving
$\Phi^\perp$ to ${\cal A}$
can be reduced to the convolution of the leading order hard 2 particle contributions with linear combination of correlators,
 thanks to the use of the Ward identity.
In the case of the 3-body vector correlator (\ref{Correlator3BodyV}), due to (\ref{epsilonRho}) the dependency of $e_T$ on
$n_\perp$ enters linearly and only through the scalar product $e^* \cdot n\,.$ Using the fact that the amplitude ${\cal A}$ depends linearly on the polarisation vector $e\,,$
 one finds that the action on the amplitude of the derivative $\frac{\partial}{\partial\,(e^*\cdot n)}$  in (\ref{nIndV}) can be extracted by the replacement $e_\alpha^* \to -p_\alpha\,.$ After using the Ward identity, it reads
\begin{eqnarray}
\label{Ward}
(y_1-y_2){\rm tr} \left[ H_{\rho}(y_1,y_2) \, p^\rho
\, \slashchar{p}\right] =
{\rm tr} \left[H(y_1) \, \slashchar{p}\right]-{\rm tr} \left[H(y_2) \, \slashchar{p}\right]\,,
\nonumber
\end{eqnarray}
as illustrated by
Fig.\ref{fig:WardIF}.
This results in the last term of r.h.s of Eq.(\ref{ninV}).
A similar treatment of
2-body correlators with transverse derivative
whose contribution can be viewed as 3-body processes with vanishing gluon  momentum leads to the l.h.s of Eq.(\ref{ninV}).
The  first term with $\varphi_1$  of the r.h.s of Eq.(\ref{ninV})
originates from the 2-body vector correlator and corresponds to the contribution of
$e_L \sim p.$ The second term with $\varphi_3$ corresponds to the contribution of the same correlator for $\rho_T$ supplemented by Eq.(\ref{epsilonRho}).
A similar treatment for axial correlators leads to Eq.(\ref{ninA}).

Since we rely on the $n-$independency of the amplitude, one may wonder about the effect of the gauge choice, which is fixed by $n\,,$ on the hard part.
The QCD Ward identities require the vanishing of the amplitude in which polarization vector of a gluon is replaced by its momentum provided all other partons are on the mass shell. In the framework of the $k_T-$factorization the $t-$channel gluons are off the mass-shell. Therefore the replacement of the $s-$channel gluon  polarization vector by its momentum leads to the vanishing of scattering amplitude up to terms proportional to $k_\perp^2/s$ where $k_\perp$ are transverse momenta of $t-$channel gluons.
From the point of view of the $t-$channel, the gauge invariance of the impact-factor means that it should vanish when the transverse momentum of any $t-$channel gluon vanishes. To acchieve this property it is necessary to include in a consistent way
not only DAs with lowest Fock state containing only quarks but also those involving
quarks and gluon, see \cite{Us}.

In practice, we here
 check this invariance by contracting the $s-$channel emitted gluon vertex in the hard part with the momentum, which
in collinear factorization is proportional to the $\rho-$meson momentum, this leads to simplifications in the use of (collinear) Ward identities. This means that the hard part is gauge invariant.

In order to prove this,
one should first project on the various color Casimir structure. In the case of the impact factor, this means to distinguish $N_c$ and $C_F$ terms. In this case, $C_F$ terms arise from
2-partons diagrams and from  3-partons diagrams where the emitted gluon
is attached to a quark line, while $N_c$ terms are obtained from  3-partons diagrams where the emitted gluon is attached to a quark line only between 2 $t-$channel exchanged gluons or from diagrams involving at least one triple gluon vertex.
Using Ward identity, the result looks like the product of the 2-parton diagrams contributing to the transition $\gamma^*_T \to \rho_L\,.$ One can readily check at Born oder that this transition vanishes, in our twist 3 treatment.
In the case of contribution proportional to $N_c$, one can prove that this
gives zero. Note that terms proportional to the virtuality $\kb^2$ remains, but they are of the form $\kb^2/s\,,$ which are higher order with respect to the
dominant $s-$power contributions to the impact factor (which scales like $s^0$) involved in the $k_T-$factorization approach. This is thus in agreement with Ward identity in perturbative Regge factorization.

Although our implementation of factorization and $n-$independence condition
was illustrated on the particular example of the impact factor at twist 3, we expect
that this procedure is more general and that the above method can be applied for other
exclusive processes, the key tool being the collinear Ward identity. This means in particular that each building block (soft and hard part, for each structure which lead to the introduction of a DA) are separately gauge invariant. This fact simplifies dramatically
the use of the $n-$independence principle.

We thus emphasize that the equations obtained above for the special case of impact factor are expected to be universal i.e.
not to refer to any specific hard process.
The equations of motion (\ref{em_rho1},\,\ref{em_rho2})
 together with
Eqs. (\ref{ninV},\,\ref{ninA}) following from $n$-independence are
four equations which constrain
seven DAs ($\varphi_1,\varphi_3,\varphi_1^T,\varphi_A,\varphi_A^T,B,D$).
Thus, any hard process involving the
exclusive production of a $\rho$ can be expressed, at twist 3, in terms
of the three independent DAs which  we choose as
$\varphi_1$, $B$ and $D$.
The solution of Eqs. (\ref{em_rho1},\,\ref{em_rho2},\,\ref{ninV},\,\ref{ninA}) with vanishing
$B$ and $D$ DAs corresponds to the Wandzura-Wilczek approximation and
takes the form \cite{GENERIC}
\begin{eqnarray}
\label{WWphi3}
\!\varphi_{3/A}^{WW}(y)&\!\!=\!\!&\frac{1}{2}\left[ \int\limits_0^y\,\frac{dv}{\bar v}\varphi_1(v)
\pm\int\limits_y^1\frac{dv}{v}\varphi_1(v)   \right]\,,\\
\label{WWphi1T}
\!\varphi_{1/A}^{T\,WW}(y)&\!\!=\!\!&\frac{1}{2}\left[
-\bar y \! \int\limits_0^y\,\frac{dv}{\bar v}\varphi_1(v)
\pm y\!\int\limits_y^1\frac{dv}{v}\varphi_1(v)   \right]\!.
\end{eqnarray}
The remaining genuine twist 3 DAs read:
\begin{eqnarray}
\label{Resphi3}
\varphi_3^{gen}(y)&=&-\frac{1}{2}\int\limits^1_{y} \frac{du}{u}
\biggl[
\int\limits^u_0 dy_2 \frac{d}{du} (\zV B-\zA D)(y_2,\, u)
\biggr.
\nonumber\\
\biggl.
&&-\int\limits^1_u \frac{dy_2}{y_2-u}  (\zV B-\zA D)(u, \,y_2)
\biggr.
\nonumber\\
\biggl.
&&-\int\limits^u_0 \frac{dy_2}{y_2-u}  (\zV B-\zA D)(y_2, \, u)
\biggr]
\nonumber \\
&&-\frac{1}{2}\int\limits^{y_1}_{0}  \frac{du}{\bar{u}}
\biggl[
\int\limits^1_u dy_2 \frac{d}{du} (\zV B+\zA D)(u, \,y_2)
\biggr.
\nonumber\\
\biggl.
&&-\int\limits^1_u \frac{dy_2}{y_2-u}  (\zV B+\zA D)(u, \,y_2)
\biggr.
\nonumber\\
\biggl.
&&-\int\limits^u_0 \frac{dy_2}{y_2-u}  (\zV B+\zA D)(y_2, \, u)
\biggr]\,,
\end{eqnarray}
\begin{equation}
\label{Resphi1T}
\varphi_1^{T \, gen}(y) =\int\limits^y_0 du \, \varphi_3^{gen}(u) - \zV \int\limits^y_0 dy_1 \int\limits^1_y dy_2 \frac{B(y_1,\, y_2)}{y_2-y_1}\,,
\end{equation}
while the corresponding expressions for $\varphi^{gen}_A(y)$ and
   $\varphi_A^{T\, gen}(y)$ are obtained by the substitutions:
\begin{eqnarray}
\label{ResphiA}
\varphi_{A}^{gen}(y) &\stackrel{\zV B \,\leftrightarrow \,\zA D}{\longleftrightarrow}& \varphi_{3}^{gen}(y)\, ,\\
\label{ResphiAT}
\varphi_{A}^{T\, gen}(y) &\stackrel{\zV B \,\leftrightarrow \,\zA D}{\longleftrightarrow}& \varphi_{1}^{T\, gen}(y)\,.
\end{eqnarray}

Let us now discuss an equivalent way for obtaining the result (\ref{Resphi1T}).
We start with the following operator equation \cite{BalBraun} written in a
gauge-invariant form,
where on the right-hand side we only keep the contributions of
vector correlators
\begin{eqnarray}
\label{Opder}
&&\frac{\partial}{\partial z_\alpha}
\biggl[
\bar\psi(z)\gamma_\mu [z,\,-z]\psi(-z)
\biggr]=
\\
&&- \bar\psi(z)\gamma_\mu [z,\,-z] \!\stackrel{\longrightarrow}
{D_{\alpha}} \psi(-z) +
\bar\psi(z)\!\stackrel{\longleftarrow}
{D_{\alpha}}\gamma_\mu [z,\,-z]\psi(-z)
\nonumber\\
&&-ig \int\limits_{-1}^{1} dv\, v\, \bar\psi(z)[z,\,vz]
z_\nu G_{\nu\alpha}(vz) \gamma_\mu [vz,\,-z]\psi(-z)\, ,\nonumber
\end{eqnarray}
where $[\,]$ denotes the Wilson line and
$\stackrel{\longrightarrow}{D_{\alpha}}=
\stackrel{\longrightarrow}{\partial_{\alpha}}-ig A_\alpha(-z)$
and
$\stackrel{\longleftarrow}{D_{\alpha}}=
\stackrel{\longleftarrow}{\partial_{\alpha}}+ig A_\alpha(z)$. Note that in (\ref{Opder}),
the derivatives
act on the arguments of the fermion fields.
In principle, the path in the Wilson line can be an arbitrary one,
provided it lies
 entirely  on the light-cone.
However, for simplicity, we choose the path to be a straight
line.
We now compute the matrix elements
of the various part of Eq.(\ref{Opder}).

First, using the parametrization (\ref{Correlator3BodyV}), the matrix element of the quark-gluon part on r.h.s of Eq.(\ref{Opder})
reads
\begin{eqnarray}
\label{Corr-qgoper}
&&\hspace{-.4cm}\langle \rho(p) | \bar\psi(z)\gamma_\mu \left\{ [z,\,-z] A_{\alpha}(-z)+ A_{\alpha}(z) [z,\,-z]\right\}\psi(-z)|0\rangle  \nonumber \\
&&\hspace{-.4cm}-\int\limits_{-1}^{1} dv\, v\,
\langle \rho(p) |\bar\psi(z)[z,\,vz]
z_\nu G_{\nu\alpha}(vz) \gamma_\mu [vz,\,-z]\psi(-z)|0\rangle
\nonumber\\
&&\hspace{-.5cm}=\!2\,m_\rho \fV\, p_\mu\, e^{*\,T}_\alpha\!\!\!
\int\limits_0^1 \!\! dy\, e^{i(y-\bar{y})p\cdot z}\!\!\!\int\limits_{0}^y \! dx_1\!\!\int\limits_y^1 \!dx_2\, \frac{B(x_1,x_2)}{x_2-x_1}\,,
\end{eqnarray}
where in the last stage integration by parts has been used.

Further, we turn on the quark-antiquark operator in (\ref{Opder}) and its vacuum--to--meson correlator. Using (\ref{epsilonRho}), the correlator (\ref{CorrelatorV})  can be  rewritten as
\begin{eqnarray}
&&\hspace{-.2cm}\langle \rho(p) | \bar\psi(z)\gamma_\mu [z,\,-z] \psi(-z)|0\rangle  \\
&&=
-2i\, m_\rho \, f_\rho \, p_{\mu} (e^*\cdot z) \int\limits_0^1 dy\, e^{i(y-\bar{y})\,p\cdot z} h(y) + \ldots \,,
\nonumber \\
&& \mbox{where}
\label{defh} \quad h(y) = \int\limits^y_0 du \, (\varphi_1(u)-\varphi_3(u))\,,
\end{eqnarray}
where the ellipsis denotes terms which are responsible for corrections
of $m^2_\rho$-order and, therefore, they are part of   twist 4 corrections.
\cite{BB}, which
 we omit in this paper.

Up to twist 3, we have
\begin{eqnarray}
\label{DerOPMatrix}
&&\frac{\partial}{\partial z_\alpha}
\biggl[\langle \rho(p) | \bar\psi(z)\gamma_\mu [z,\,-z] \psi(-z)|0\rangle \biggr]\nonumber \\
&&=
-2i\, m_\rho \, f_\rho \,p_{\mu} e^*_\alpha \int\limits_0^1 dy\, e^{i(y-\bar{y})\,p\cdot z} h(y)\, .
\end{eqnarray}
Finally, the pure quark part of r.h.s of Eq.(\ref{Opder}) together with the
parametrization (\ref{CorrelatorDerV}) and the translation invariance  yields,
\begin{eqnarray}
\label{MatrixDerTrans}
\langle \rho(p)|
\bar\psi(z)\gamma_{\mu}
\left[\stackrel{\longrightarrow}{\partial^\perp_{\alpha}} -
\stackrel{\longleftarrow} {\partial^\perp_{\alpha}}
\right]
\psi(-z) |0\rangle
\nonumber  \\
= -2i \,m_\rho \, f_\rho \,p_{\mu} e^{*\, T}_\alpha \int\limits_0^1 dy\, e^{i(y-\bar{y})\,p\cdot z}
\varphi_1^T(y)\, .
\end{eqnarray}
Thus, using (\ref{Corr-qgoper},\,\ref{DerOPMatrix},\,\ref{MatrixDerTrans}),
the matrix element of Eq.(\ref{Opder}) leads to
\begin{eqnarray}
 \label{h_reduced}
h(y) =
- \varphi_1^T(y) - \zV \int\limits_{0}^y dx_1\int\limits_y^1 dx_2\, \frac{ B(x_1,x_2)}{x_2-x_1}
\end{eqnarray}
which reproduces the results (\ref{Resphi3}, \ref{WWphi3}, \ref{WWphi1T})
obtained previously. This can be seen most easily by inserting in Eq.(\ref{h_reduced})
the expressions (\ref{Resphi1T}) for $\varphi_1^{T \, gen}$ and (\ref{WWphi1T})
for $\varphi_1^{T \, WW}\,.$ The $B$ terms then cancel out and what remains to be shown is that
the r.h.s  of (\ref{defh}) is reproduced. The $\varphi_3^{gen}$ is readily obtained
and the integral of $\varphi_1$ and of $\varphi_3^{WW}$ in the r.h.s  of (\ref{defh}) is correctly reproduced after interchange of integrations.
This operator approach can be applied similarly to the axial
part.

We now relate our results
(\ref{Resphi3}, \ref{ResphiA}),
 with those obtained in studies of DAs based on the conformal
expansion on the (coordinate space) light cone \cite{BB}. Comparison between  (\ref{Correlator3BodyV}, \ref{Correlator3BodyA}) and similar expressions of Ref.\cite{BB} in the light-cone $n \cdot A=0$ gauge leads  to the
 identification
\begin{eqnarray}
\label{DictBD}
 B(y_1,\,y_2)&=&-\frac{1}{m_\rho}\frac{V(y_1, \, 1-y_2, \, y_2-y_1)}{y_2-y_1} \\
D(y_1,\,y_2)&=&-\frac{1}{m_\rho}\frac{A(y_1, \, 1-y_2, \, y_2-y_1)}{y_2-y_1}\,.
\end{eqnarray}
Using these expressions for $V$ and $A$ one can now recover our results (\ref{Resphi3}, \ref{ResphiA}) after straighforward but tedious manipulations, provided
the important identification is done, for the vector projection:
\begin{eqnarray}
\label{relBBvector}
&&\varphi_1(y)=
\phi_{\parallel}(y) ,
\quad
\varphi_3(y)=
 g^{(v)}(y) \,,
\end{eqnarray}
and for the axial projection
\begin{eqnarray}
\label{relBBaxial}
&&\varphi_A(y) =
-\frac{1}{4} \frac{\partial g^{(a)}(y)}{\partial y}\,.
\end{eqnarray}
In summary, we describe consistently exclusive
processes at higher twist.  The specific
$\gamma^* \to \rho$ impact factor will indeed be shown to be gauge invariant
and free from end-point singularities in  a separate publication \cite{Us}.
The $n$-independency condition generalized up to the dynamical
twist $3$ plays a crucial role for
the consistency of this approach with the studies of DAs performed in
 \cite{BB}.

The present framework opens the way to a systematic and consistent treatment of
hard exclusive  processes.
This does not preclude the solution of the well known
end-point singularity problem \cite{MP, GK} which may still require a separate treatment.
Let us finally emphasize
the simplicity of the present method to perform practical
calculations \cite{Us}.


\acknowledgments
\noindent
We thank O.~V.~Teryaev for discussions.
 This work is partly supported by the ECO-NET program, contract
18853PJ, the French-Polish scientific agreement Polonium, the grant
ANR-06-JCJC-0084, the RFBR (grants 09-02-01149,
 08-02-00334, 08-02-00896) and
the Polish Grant N202 249235.


\begin{thebibliography}{99}
\bibitem{Nind} Note that this introduces a non-physical $n-$dependence which we will discuss below.
\bibitem{RG} The appearance of the total  derivative may be interpreted as a (vector) analog to the renormalization group (RG) invariance when the
dependences on the renormalization parameter coming from various sources
 cancel.  One can view this as a RG-like flow in the space of light-cone
directions of contributions to the amplitude where the polarization vector plays the
 role of a beta function.
\bibitem{GENERIC} We use the generic notation $\varphi=\varphi^{WW}+\varphi^{gen}\,.$
\end{thebibliography}
\end{document}